\documentclass[sigconf,screen]{acmart}

\acmConference{}{}{}
\setcopyright{none}
\acmDOI{}
\acmISBN{}
\acmYear{}
\acmSubmissionID{ase26ind-p2-p}
\received{2026-04-21}
\received[accepted]{2026-07-01}

\usepackage{booktabs}
\usepackage{balance}
\usepackage{microtype}

\begin{document}
% \raggedbottom

\title{DragonCrawl: A Generative, Intent-Based Framework for Scalable Mobile End-to-End Testing}

%% Authors
\author{Sowjanya Puligadda}
\correspondingauthor
\orcid{0009-0003-9087-1212}
\authornote{Main authors. Contributed equally.}
\affiliation{%
  \institution{Uber Technologies, Inc.}
  \city{San Francisco}
  \state{CA}
  \country{USA}
}
\email{sowjanyap@uber.com}

\author{Mengdie Zhang}
\orcid{0009-0009-3149-1643}
\authornotemark[1]
\affiliation{%
  \institution{Uber Technologies, Inc.}
  \city{San Francisco}
  \state{CA}
  \country{USA}
}
\email{mengdiez@uber.com}

\author{Ali Zamani}
\orcid{0000-0001-5846-0374}
\authornotemark[1]
\affiliation{%
  \institution{Uber Technologies, Inc.}
  \city{San Francisco}
  \state{CA}
  \country{USA}
}
\email{azamani@uber.com}

\author{Dhruva Dixith Kurra}
\orcid{0009-0004-2636-8134}
\affiliation{%
  \institution{Uber Technologies, Inc.}
  \city{San Francisco}
  \state{CA}
  \country{USA}
}
\email{dkurra@uber.com}

\author{Eric Chen}
\orcid{0009-0000-6932-8248}
\authornote{Senior Advisors.}
\affiliation{%
  \institution{Uber Technologies, Inc.}
  \city{San Francisco}
  \state{CA}
  \country{USA}
}
\email{echen@uber.com}

\author{Juan Marcano}
\orcid{0009-0005-9363-1489}
\authornotemark[2]
\affiliation{%
  \institution{Uber Technologies, Inc.}
  \city{San Francisco}
  \state{CA}
  \country{USA}
}
\email{marcano@uber.com}

\renewcommand{\shortauthors}{Puligadda et al.}

\begin{abstract}
As mobile applications grow in complexity, traditional End-to-End (E2E) testing frameworks struggle with UI volatility, maintenance overhead, and cross-platform scalability. This paper presents DragonCrawl, an AI-driven mobile testing system for continuous regression testing that has evolved from embedding-based similarity matching to generative intent-based reasoning using large language models. Unlike prior LLM-based testing research focused on exploratory testing and crash detection, DragonCrawl validates specific user flows on every code change, blocking commits that break critical functionality. By leveraging GPT-4o's multimodal capabilities, DragonCrawl achieves 91.6\% pass rate on iOS and 92.2\% on Android across 1,013 automated tests running continuously in CI/CD pipelines. The system reduces test onboarding time from 96-120 hours to under 4 hours and has saved an estimated 27 developer years in test maintenance effort. We present the architectural evolution from V1 (semantic embedding matching) to V2 (generative intent-based reasoning), discuss implementation challenges including token explosion and memory constraints, and report operational experience from production deployment. The integration of multimodal vision for end-state detection and tool calling for backend state transitions enables comprehensive regression testing that bridges UI interactions with system state. Our results demonstrate that AI-driven testing can maintain stability while eliminating the brittleness of traditional automated tests, enabling continuous quality assurance at scale.
\end{abstract}

%% CCS concepts and keywords (required by acmart)



\maketitle

\begin{center}
\footnotesize
\textbf{Author-accepted manuscript.} Accepted at ASE-Industry 2026.
The final version will appear in the ACM Digital Library.
\end{center}

\section{Introduction}

Mobile applications must function correctly in diverse environments: languages, cities, device types, and operating systems. At Uber, applications serve 200 million monthly active users in 15,000+ cities in 60+ languages. Each of 61 critical flows must work across all these contexts, creating a combinatorial explosion: 61 flows $\times$ 150 cities $\times$ 60 languages = 549,000 test cases.

Manual testing cannot scale. Testing 61 flows in 150 cities requires thousands of manual executions, representing weeks of effort. Automated UI testing addresses scalability, but suffers from brittleness: tests rely on hardcoded element locators that break with UI changes. Test maintenance consumes over 30\% of testing effort \cite{coppola2020}.

Recent LLM-based testing research shows promise but focuses on exploratory testing and crash detection. Tools like GPTDroid~\cite{gptdroid} and DroidAgent~\cite{droidagent} measure success through activity coverage and bugs discovered, running for fixed periods to explore applications. These excel at finding unknown issues but cannot answer the CI/CD question: "Does this code change break our critical flows?"

DragonCrawl addresses continuous regression testing at production scale. Unlike exploratory tools, it validates specific flows on every code change, blocking commits that break functionality. The system evolved from embedding-based similarity matching (V1, 2023) to generative intent-based reasoning (V2, 2024). V1 achieved 80-82\% pass rates on simple flows but 0\% on complex flows. V2 improvements:

\begin{itemize}
\item Pass rates: 80--82\% (V1) to 91.6--92.2\% (V2)
\item Onboarding: 96--120 hours to $<$4 hours
\item Coverage: 48 to 1,013 tests
\item Saved: 27 developer years in maintenance and authorship
\end{itemize}

This paper contributes:

\begin{itemize}
\item Shift from exploratory to regression testing: validates specific flows on every code change in CI/CD pipelines
\item Intent-based architecture scaling across languages, markets, and UI variations
\item Multimodal end-state detection using vision
\item Tool calling for backend state transitions
\item Evaluation across 1,013 production tests
\item Operational experience including organizational transformation and cost optimization
\end{itemize}

The remainder of this paper is organized as follows. Section~\ref{sec:background} provides background on mobile testing challenges and motivates the need for AI-driven approaches. Section~\ref{sec:architecture} presents the DragonCrawl V2 architecture. Section~\ref{sec:implementation} discusses implementation and scalability challenges. Section~\ref{sec:evaluation} evaluates performance in multiple dimensions. Section~\ref{sec:operations} reports operational experience and organizational impact. Section~\ref{sec:related} surveys related work. Section~\ref{sec:limitations} acknowledges limitations. Section~\ref{sec:conclusion} concludes with lessons learned and future directions.

\section{Background and Motivation}
\label{sec:background}

\subsection{The Mobile Testing Challenge}

At Uber, mobile applications must function correctly across 15,000+ cities in 60+ languages, serving 200 million monthly active users. Each of 61 critical flows must work across all markets, languages, and device configurations. Manual testing cannot scale: testing 61 flows across 150 cities requires thousands of executions representing weeks of effort per cycle. Automated UI testing addresses scalability but suffers from brittleness: hardcoded element locators break with any UI change. Research shows test maintenance consumes over 30\% of mobile testing effort \cite{coppola2020}.

The problem compounds through combinatorial explosion. Testing 61 flows $\times$ 150 cities $\times$ 60 languages yields 549,000 test cases. Each dimension multiplies requirements exponentially. Traditional automated testing cannot scale to this coverage: writing and maintaining tens of thousands of tests is impractical, and manually executing them would take weeks.

\subsection{DragonCrawl V1: Embedding-Based}

DragonCrawl V1 (2023) used sentence embeddings (MPNet~\cite{mpnet}) and K-D tree similarity matching against a "Golden Dataset" of reference screen states. While V1 achieved 80-82\% pass rates on simple flows like UberX trip booking, it faced critical limitations.

\textbf{High Onboarding Costs}: Building the Golden Dataset required 4-5 days per flow. Engineers manually captured reference screens, labeled actions, and validated sequences. The UberX flow alone required 130+ annotated reference states, making scaling to dozens of flows impractical.

\textbf{Token Explosion and Memory Constraints}: MPNet's limited context window caused truncation on complex screens with many UI elements. More critically, V1 lacked action history. When identical screens appeared at different flow points (e.g., home screen at start vs. after task completion), the similarity model could not distinguish between them, causing infinite loops.

These technical limitations culminated in complex flow failures. Flows requiring repeated actions—like Uber Pool, which requires selecting multiple pickup points in sequence—confused V1 because each selection step looked similar. Pass rates dropped below 5\% on these scenarios, making V1 unsuitable for 2024 and beyond as the team needed to scale to more complex production tests.

The transition to V2 with GPT-4o introduced costs, but token prices were declining 10$\times$ every 6-12 months. Combined with prompt caching, costs became manageable relative to the engineering time V1's limitations were consuming.

\subsection{The Promise of Generative AI}

Recent advances in large language models have demonstrated remarkable capabilities in natural language understanding, reasoning, and planning. Models like GPT-4 can interpret complex instructions, maintain context over long conversations, and generate coherent action sequences to achieve specified goals. These capabilities map directly onto the requirements for mobile testing.

Generative models offer several advantages over embedding-based approaches:

\textbf{Full Context Windows}: Modern LLMs can process much longer contexts (128K+ tokens for GPT-4). This eliminates token explosion problems, allowing the system to reason about complex screens and maintain long action histories.

\textbf{True Reasoning}: Rather than finding similar reference states, generative models can reason about what action makes sense given the current screen, the testing goal, and the action history. This enables handling novel situations not seen in training data.

\textbf{Multimodal Understanding}: Models like GPT-4o process both text and images. This allows visual verification of test outcomes, mimicking how human QA engineers validate results by looking at screenshots.

\textbf{Tool Calling}: LLMs can be equipped with tools to perform external actions. This enables bridging the gap between UI testing and backend state management, solving problems where tests are blocked by database states or external dependencies.

The transition to V2 represents a paradigm shift from pattern matching to genuine understanding. Instead of asking "what reference state is most similar to this screen?", the system now asks "given my goal, what I see on screen, and what I've done so far, what should I do next?" This change enables handling the complexity and variability of production mobile applications at scale.

\section{DragonCrawl V2 Architecture}
\label{sec:architecture}

The V2 architecture decouples navigation from static matching, replacing the similarity-based approach with generative reasoning. The system maintains full action history, reasons about current screen state in context of testing goals, and uses multimodal vision to validate outcomes.

DragonCrawl's intent-based model represents a fundamental shift in how we approach mobile end-to-end (E2E) automation. Previously, DragonCrawl relied on a text embedding retrieval system that matched the current UI screen to the most semantically similar historical screen and replayed its associated action. While this approach enabled early success, such as running the core trip flow and blocking P0 bugs, it was inherently limited by static similarity search. It lacked awareness of flow progression, action history, and high-level goals, which made it brittle as we expanded to more complex and diverse user journeys.

\subsection{Intent-Based Reasoning Engine}

The intent-based model reframes mobile testing as a goal-directed reasoning problem rather than a nearest-neighbor lookup task. Instead of asking "Which past screen looks most similar to this one?", we now ask "Given the overall intent, defined steps, current UI state, and past actions, what is the best next action to achieve the goal?" By leveraging GPT-4o~\cite{gpt4o} through a dedicated LangChain service, DragonCrawl can reason holistically about where it is in a flow, what has already been done, and what should logically happen next. This shift introduces path determinism, reduces ambiguity between visually similar screens, and aligns action selection more closely with how a real user would navigate the app.

We moved to this approach primarily to address three structural bottlenecks: path indeterminacy, fragile end-state detection, and unsustainable dataset maintenance. The embedding model treated each screen independently, which caused failures in flows with repeated or near-identical UI states. End-of-flow detection relied on hardcoded resource identifiers that frequently broke with UI updates. Additionally, onboarding a single flow required days of manual data collection and continuous maintenance across platforms. These constraints made scaling to all 61 critical flows and beyond operationally expensive and difficult to sustain.

The intent-based model directly solves these issues by introducing context-aware reasoning and multimodal end-state assertions. Users now define flows in natural language (i.e. intent, steps, and assertions) without manual data recording or embedding generation. End-state detection is handled via screenshot-based multimodal evaluation, making it more resilient to UI changes. As a result, DragonCrawl becomes more accurate, robust, and scalable, enabling reliable pre-land and post-land validation across platforms while significantly reducing onboarding effort and long-term maintenance overhead.

\begin{figure}[htbp]
\centering
\includegraphics[width=0.95\columnwidth]{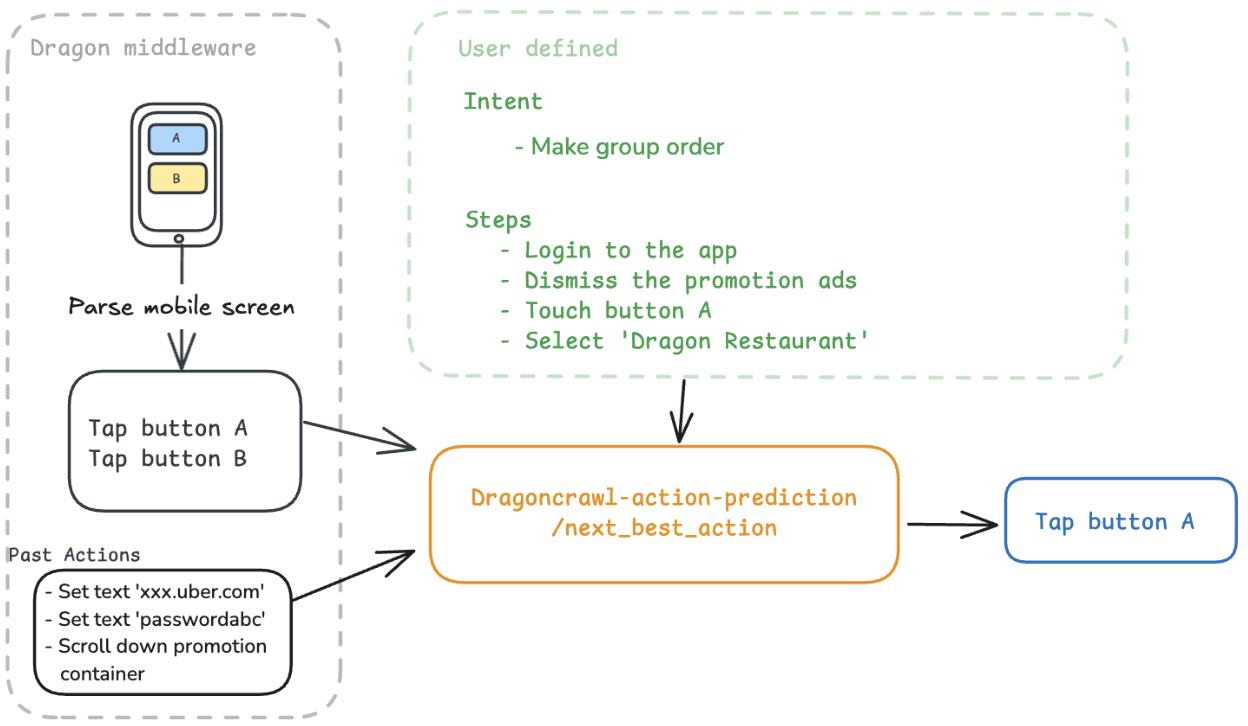}
\Description{High-level architecture diagram showing the components of the DragonCrawl system.}
\caption{Architecture of DragonCrawl.}
\label{fig:dc_arch}
\end{figure}

The core of DragonCrawl V2 is an intent-based reasoning engine that generates next actions based on rich context. For each decision point during test execution, the system constructs a context package containing:

\begin{itemize}
\item \textbf{Current UI Snapshot}: A textual representation of the XML view hierarchy of the screen, including all interactive elements, their properties, and their relationships
\item \textbf{Overall Intent}: The high-level goal specified in the test definition (e.g., "Complete a trip booking from downtown San Francisco to the airport")
\item \textbf{Action History}: A chronological log of all actions taken so far, including which elements were interacted with and what UI changes resulted
\item \textbf{Available Actions}: A list of all currently interactable elements on screen with their properties and suggested interaction types
\end{itemize}

This context package is sent to a dedicated LangChain service that interfaces with GPT-4o. The service maintains the conversation history across the test execution, ensuring the LLM has full context of the testing session.

\subsection{Multimodal End-State Detection}

End state detection is a critical component of DragonCrawl because it determines whether a mobile flow has successfully reached its intended goal. In the previous embedding-based approach, this was handled through hardcoded UI identifiers such as resource IDs or specific text strings that were manually selected for each flow and platform. While this method worked initially, it was fragile and labor-intensive. UI identifiers frequently changed due to refactors, redesigns, or platform differences between Android and iOS, causing false failures or false passes. Every flow required manual inspection to identify a "unique" element, and ongoing maintenance was needed whenever the UI evolved.

In the intent-based model, we reframe end state detection as a reasoning problem rather than a static rule-matching task. Instead of checking for specific resource IDs, we send a base64-encoded screenshot of the current UI along with a natural language assertion (e.g., "Is the ride successfully requested?") to the /assert\_flow\_end endpoint. Backed by GPT-4o's multimodal capabilities, the model evaluates the visual context together with the textual expectation and determines whether the goal has been achieved. This allows the system to reason over layout, visual hierarchy, and contextual cues, not just specific identifiers.

This approach significantly improves robustness. Screenshots capture the full visual state of the app, including elements that may not be reliably represented in XML or may vary across platforms. The model can tolerate minor UI shifts, wording changes (e.g., "Confirm" vs. "Okay"), or layout adjustments, as long as the overall intent of the screen matches the expected outcome. As a result, end state assertions become more resilient to UI evolution and require far less manual maintenance compared to identifier-based matching.

While image-based assertions introduce higher inference cost and require careful handling of transient states (e.g. loading screens), the gain in reliability and maintainability outweighs these trade-offs. Accurate end state detection is foundational to trustworthy E2E testing. By leveraging multimodal reasoning, the intent-based model ensures that DragonCrawl can confidently determine when a flow is truly complete, enabling more reliable pre-land and post-land validation at scale.

\begin{figure}[htbp]
\centering
\includegraphics[width=0.95\columnwidth]{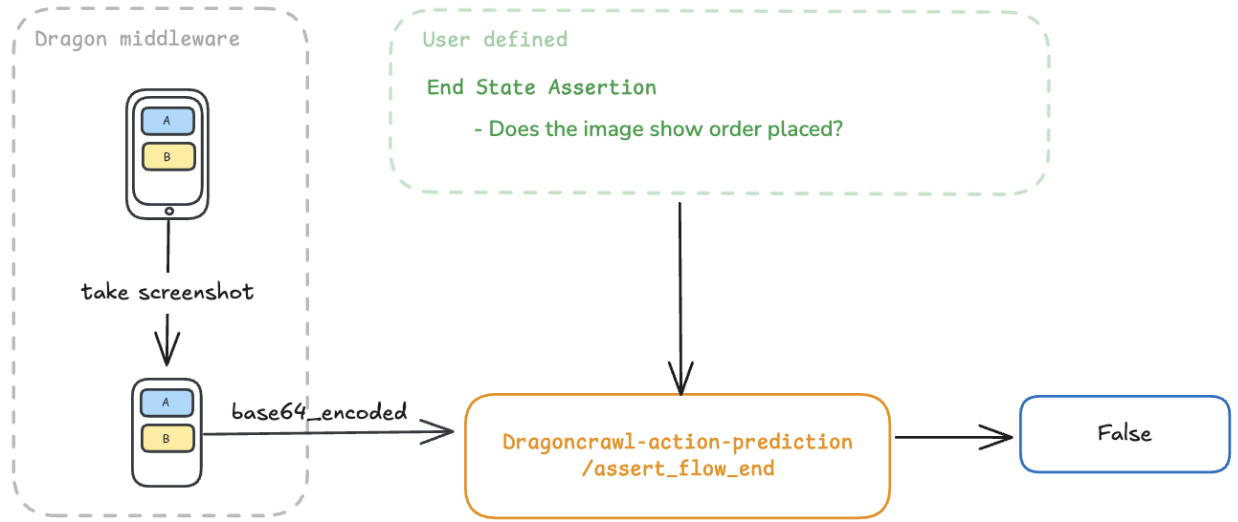}
\Description{Diagram of the dataflow for visual assertions, showing how screenshots and natural language prompts are combined and sent to the multimodal model for end-state detection.}
\caption{Dataflow of Visual Assertions.}
\label{fig:monthly}
\end{figure}

Traditional automated tests verify outcomes by asserting that specific elements exist at specific locations. This approach breaks when layouts change. DragonCrawl V2 uses multimodal vision to detect test completion.

At each step, the system captures a screenshot and encodes it as base64. When the model believes the test goal may be achieved, it triggers an end-state detection query. The system sends both the screenshot and a natural language question to GPT-4o: "Based on this image, has the user successfully reached the receipt or confirmation page showing trip details?"

The model analyzes the visual content and responds yes/no with an explanation. This mimics how human QA testers verify results: by looking at the screen and assessing whether it matches expected outcomes. The approach is robust to UI changes because it evaluates visual appearance rather than specific element properties.

Mid-state assertions work similarly. Tests can specify conditions that should hold at intermediate points: "verify that a discount code is applied", "confirm that the selected vehicle type is displayed". The system asks these as visual questions throughout execution, providing continuous validation rather than only checking final outcomes.

\subsection{Prompt Engineering Strategy}

The prompt design uses constraint-based zero-shot planning. Rather than providing few-shot examples, which would require maintaining example sets for different flow types, we specify constraints that guide the model's reasoning process.

The system prompt instructs the model to:

\begin{enumerate}
\item Analyze the current screen in relation to the testing goal
\item Review the action history to understand what has already been accomplished
\item Identify which available actions would progress toward the goal
\item Select the single best next action based on this analysis
\item Format the response as structured JSON specifying the action type and target element
\end{enumerate}

Constraints enforce safety and determinism:

\begin{itemize}
\item The model must select from the provided list of available actions and send a wait action for the screen to update
\item Responses must be valid JSON with required fields
\item Actions must reference elements by provided identifiers
\item The model should not hallucinate elements or actions not present on screen
\end{itemize}

This approach balances flexibility with control. The model has freedom to reason about the best path forward, but cannot take invalid actions or generate malformed responses.

\subsection{Context Enhancement}

To give the model contextual information about the current app state, we use textual representation. Specifically, we derive this representation from the mobile view hierarchy, which exposes structured information about UI elements, including their identifiers, visibility and interaction properties.

However, raw view hierarchy contains substantial noise. They include deeply nested layout containers, non-interactive elements, invisible nodes, and overlapping layers that do not meaningfully contribute to action selection. Directly supplying this unfiltered information to the model inflates context size and introduces ambiguity.

\begin{figure}[htbp]
\centering
\includegraphics[width=0.999\columnwidth]{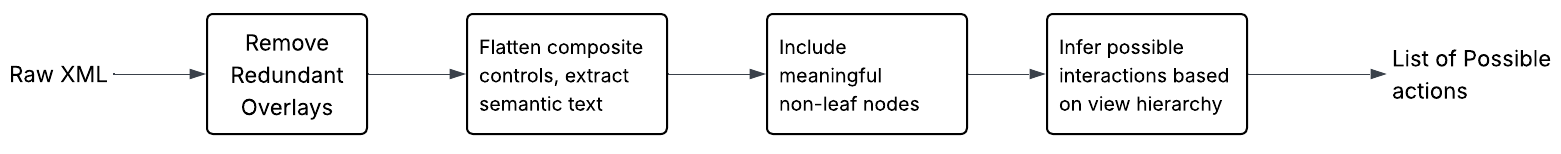}
\Description{Diagram showing the transformation from a raw Android view hierarchy with 144 nodes into a compact set of 51 executable actions used by the model.}
\caption{Context enhancement from view hierarchy.}
\label{fig:xml_flow}
\end{figure}

Therefore, we introduce a canonization process that extracts only interaction-relevant information, producing a compact and stable decision space for the model, which can be seen in Figure~\ref{fig:xml_flow}.

Thus, a raw Android view hierarchy containing a total 144 nodes (excluding the root node) and 44 leaf nodes gets transformed into 51 executable actions, including tap, scroll, input.

\subsection{Tool Calling for Backend Integration}

Many E2E tests involve backend state transitions that, while technically achievable through UI interactions, are impractical to reproduce reliably during automated execution. Consider the Uber driver onboarding process. A new driver uploads required documents through the mobile app and waits for approval before being allowed to go online. While document upload can be performed via UI automation, the approval process itself is unsuitable for regression testing at scale.

To ensure deterministic and efficient test execution, we move from state-matching to goal-conditioned, cross-layer decision policies. We introduce a unified reasoning loop that interleaves UI interaction and backend state orchestration during runtime.

Instead of waiting for real approval, the system can trigger backend API calls in controlled test environments to simulate document approval. Once the backend state is updated, the system proceeds to validate that the UI correctly reflects the driver's updated eligibility status.

The workflow operates as follows:

\begin{enumerate}
\item The model identifies that a mid-state assertion or prerequisite action requires backend changes
\item It generates a tool call request specifying the API endpoint, parameters, and expected effect
\item The LangChain service executes the gRPC/HTTP call against backend test environments
\item The result (success/failure with details) is fed back to the model
\item The model decides the next UI action based on the backend operation result
\end{enumerate}

Tool calls in DragonCrawl V2 are dynamically generated based on runtime context. Rather than following predefined scripts, backend requests are conditioned on:

\begin{itemize}
\item Current test account state
\item The user's position within the test flow
\item Prior backend responses
\item Intermediate UI observations
\end{itemize}

\begin{figure}[htbp]
\centering
\includegraphics[width=0.95\columnwidth]{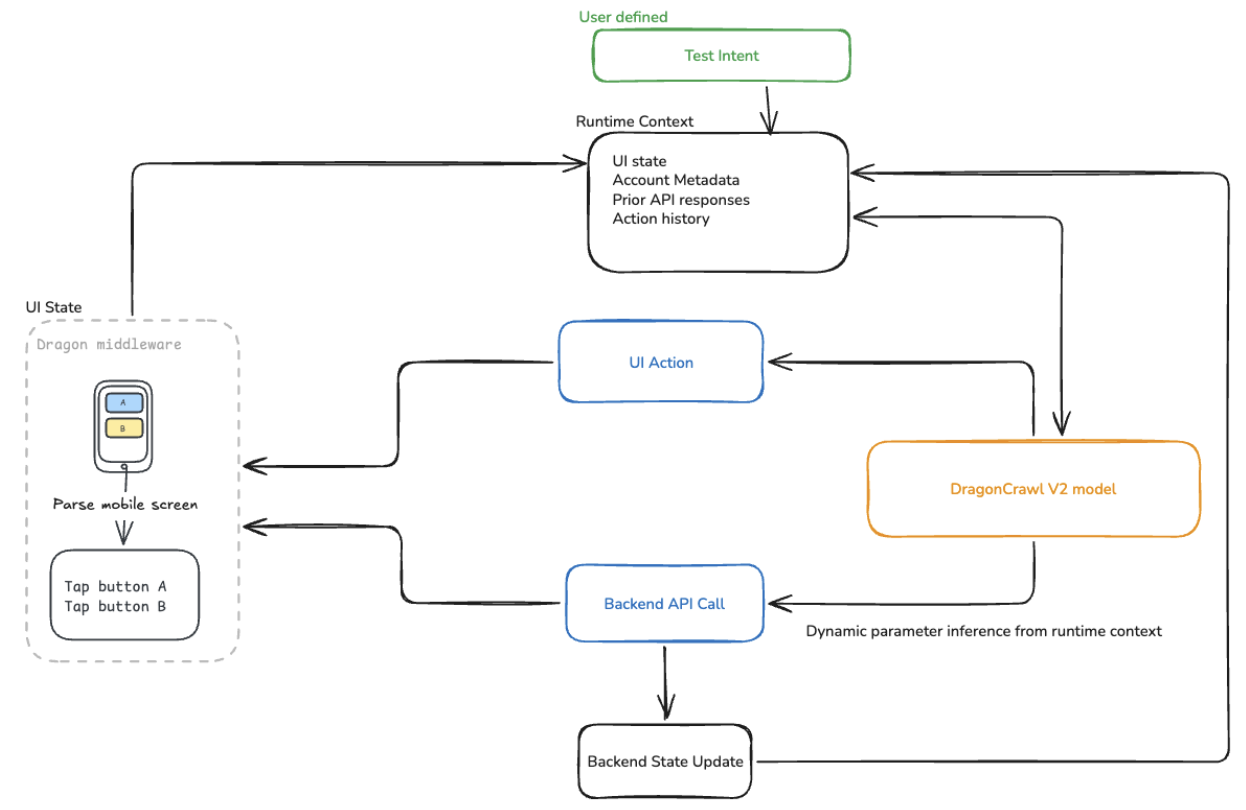}
\Description{Diagram illustrating how DragonCrawl conditions backend tool calls on current test state, flow position, prior backend responses, and UI observations.}
\caption{Context-Conditioned Backend Orchestration in DragonCrawl.}
\label{fig:backend_context}
\end{figure}

This allows backend operations to adapt to evolving test conditions, transforming backend state transitions into actions within the model's decision policy. This bridges the gap between UI testing and system state management. Tests can now set up complex scenarios, trigger edge cases by manipulating backend state, and validate that the mobile UI responds correctly to backend changes.

\subsection{Security and Access Control}

Integrating LLMs with production systems introduces security considerations. DragonCrawl implements several protective mechanisms:

\textbf{Data Masking}: Only non-personal data is used in tests. PII redactors scan all input before sending to the LLM, masking phone numbers, email addresses, and other sensitive information.

\textbf{Template-Based Tool Calling}: Backend API templates are stored in restricted blob storage. Tests reference templates by name rather than specifying raw API calls. This prevents prompt injection attacks that might try to execute arbitrary API calls.

\textbf{Service Proxy Architecture}: The LangChain service uses the existing authentication infrastructure at Uber. CI environments do not have direct access to backend APIs or LLM endpoints. All requests route through the authenticated proxy, which validates permissions and enforces rate limits.

\textbf{Audit Logging}: All LLM interactions are logged to Hive tables with full context, decisions, and outcomes. This provides traceability for debugging and security monitoring.

\section{Implementation and Scalability}
\label{sec:implementation}

Deploying an LLM-based testing system at production scale required significant engineering across the CI/CD pipeline, evaluation infrastructure, and cost optimization.

\subsection{CI/CD Pipeline Integration}

The DragonCrawl execution flow integrates with Uber's Buildkite CI infrastructure:

\begin{enumerate}
\item \textbf{Environment Setup}: Provision Android/iOS emulators or physical device farms with required app versions
\item \textbf{Context Construction}: Read intent files from the test case management system containing test goals, mid-state assertions, and tool call templates
\item \textbf{Inference Loop}: Execute continuous calls to LangChain endpoints:
  \begin{itemize}
  \item \texttt{/next\_best\_action}: Get next action recommendation
  \item \texttt{/assert\_flow\_end}: Check if test goal is achieved
  \item \texttt{/verify\_mid\_state}: Validate intermediate conditions
  \end{itemize}
\item \textbf{Metric Emission}: Stream results via Kafka to Grafana dashboards tracking pass rates, execution times, and failure modes
\end{enumerate}

The pipeline supports parallel execution across device configurations. A single test definition runs simultaneously on multiple iOS and Android versions, different screen sizes, and various locale settings. Results aggregate to provide coverage across the configuration matrix.

\subsection{Golden Dataset and Evaluation}

To prevent hallucination and ensure consistent behavior, DragonCrawl maintains a Golden Dataset of approximately 10,000 validated interaction sequences. Every test execution logs its complete trace (screens encountered, actions taken, outcomes achieved) to Hive tables.

The Michelangelo ML Studio runs daily evaluation pipelines:

\begin{enumerate}
\item Sample recent test executions from production
\item Compare action sequences against Golden Dataset for known flows
\item Flag divergences where the model took different actions than expected
\item Compute accuracy metrics: action correctness, goal achievement rate, assertion pass rate
\item Alert if accuracy drops below 95\% threshold
\end{enumerate}

This continuous evaluation catches model degradation. If prompt changes, model updates, or UI evolution cause accuracy drops, the system alerts engineers before impact spreads to many tests.

When the daily evaluation flags a divergence, the follow-up is human, but not on every case. Flagged divergences accumulate and are triaged on a roughly weekly cadence, and we deduplicate before review so engineers are not asked to judge the same case repeatedly. A divergence that recurs is reviewed; one that appears only once is treated as a one-off, typically a transient pop-up or device hiccup that did not reproduce, and is not escalated. This effort is unevenly distributed: the 61 core flows rarely require review, while lower-priority flows generate most of it, especially those tied to highly active, rapidly changing features. Recurring cases fall into two kinds: legitimate improvements, where the model found a better path, and genuine errors, which prompt refinement or added constraints. Because most runs never diverge and one-offs are filtered out, manual effort stays small: reviewers spend roughly 5\% of their time on flow maintenance and review combined, about a quarter of it on this weekly triage. This keeps human attention on the cases the system cannot resolve on its own while still surfacing novel interaction patterns.

% Human reviewers periodically validate flagged divergences. Some represent legitimate improvements where the model found a better path. Others indicate errors that require prompt refinement or additional constraints. This human-in-the-loop process maintains quality while allowing the system to discover novel interaction patterns.

LLM inference is expensive, but workable. Early estimates suggested $\$1.8M$ annual costs at Uber's testing scale. Aggressive optimization reduced projected costs to approximately $\$200K$ annually. Several strategies achieved this reduction:

\textbf{Prompt Caching}: GPT-4o supports prompt caching where repeated context is cached on the server. DragonCrawl exploits this by structuring prompts to maximize cache hits. The system prompt, test intent, and mid-state assertions remain constant across multiple action decisions within a test execution. Only the current screen state and action history change. This achieves 80\%+ cache hit rates, dramatically reducing token processing costs.

\textbf{Selective Execution}: Not all code changes require full test suite execution. The system integrates with Uber's diff analysis pipeline to identify which tests are relevant to code changes. Only tests for affected features run on pre-land validation. Full suites run nightly.

\textbf{Reduced Inference Volume}: V1 required separate inference calls for every screen in the Golden Dataset to compute similarity scores. V2 makes one inference call per action decision, plus occasional calls for end-state and mid-state assertions. For typical flows, this reduces total inference volume by 60\%.

\textbf{Batch Processing}: Assertions for multiple mid-state conditions can be evaluated in a single LLM call by asking multiple visual questions together. This amortizes per-request overhead.

\subsection{Handling Model Errors}

LLMs occasionally produce unexpected outputs. DragonCrawl implements several error handling mechanisms:

\textbf{Response Validation}: All LLM responses are validated against expected schemas before execution. Malformed JSON, missing required fields, or references to non-existent elements are caught and trigger retries with additional constraints.

\textbf{Action Verification}: Before executing actions, the system verifies that target elements still exist and are interactable. UI can change between when a screenshot was taken and when an action executes. If an element is no longer available, the system captures a new screenshot and requests a new action.

\textbf{Infinite Loop Detection}: If the system executes the same action repeatedly without UI state changes, it triggers an escape mechanism that asks the model to try a different approach or declare the test blocked.

\textbf{Timeout Handling}: Each action decision has a timeout. If the LLM does not respond within the threshold, the system fails gracefully rather than hanging indefinitely.

These mechanisms maintain test reliability despite occasional model errors. In practice, less than 0.5\% of test executions fail due to LLM issues after implementing these safeguards.

\section{Evaluation}
\label{sec:evaluation}

We evaluate DragonCrawl V2 across multiple technical dimensions: prompting strategy selection, model performance on action prediction, pass rate improvements, execution latency, and cost optimization through caching.

\subsection{Model Performance: Action Prediction}

We evaluate the model's ability to predict the correct next action using precision@k, measuring whether the correct action appears in the model's top-k predictions. Table~\ref{tab:precision} shows results.

\begin{table}[htbp]
\centering
\caption{Action Prediction Precision@k for DragonCrawl V2.}
\label{tab:precision}
\setlength{\tabcolsep}{4pt}% default is 6pt
\begin{tabular}{lccc}
\toprule
\textbf{Metric} & \textbf{Precision@1} & \textbf{Precision@2} & \textbf{Precision@3} \\
\midrule
Simple flows & 96.2\% & 98.5\% & 99.1\% \\
Medium flows & 91.8\% & 95.7\% & 97.3\% \\
Complex flows & 87.4\% & 93.1\% & 95.8\% \\
\midrule
\textbf{Overall} & \textbf{92.1\%} & \textbf{96.0\%} & \textbf{97.5\%} \\
\bottomrule
\end{tabular}
\end{table}

The model achieves 92.1\% precision@1, meaning it selects the correct action on the first attempt in 92\% of decisions. For the remaining cases, the correct action appears in the top-3 predictions 97.5\% of the time. This high precision is critical for test reliability, as incorrect actions can lead to test failures or infinite loops.

Performance varies by flow complexity. Simple flows with clear navigation paths achieve 96.2\% precision@1. Complex flows with conditional branching and repeated actions achieve 87.4\% precision@1, still substantially better than V1's embedding-based approach which achieved below 50\% on similar flows.

\subsection{Pass Rate Analysis}
\label{sec:pass_rate_analysis}

A test \emph{passes} when DragonCrawl reaches the flow's defined end state (e.g. the trip rating screen) and every visual assertion specified by the test owner is satisfied (e.g., ``does the screen show the driver's license plate?''). The verdict is system-determined, with the end state and each true/false assertion evaluated from the captured screens. The underlying visual question answering is highly reliable, with near-perfect accuracy reported in our prior work~\cite{chaostesting}, so blocking is governed by policy rather than model error: a commit is blocked when the end state is not reached or a high-priority assertion fails, while low-priority assertion mismatches (e.g. the text in a button changing from \emph{"confirm"} to \emph{"ok"})  are tracked but non-blocking, since gating deployment on negligible-impact issues is not operationally worthwhile.

Table~\ref{tab:pass_rates} shows end-to-end pass rates across DragonCrawl versions. V2 achieves 91.6\% on iOS and 92.2\% on Android, compared to 80--82\% for V1.

\begin{table}[htbp]
\centering
\caption{End-to-End Pass Rates Across DragonCrawl Versions.}
\label{tab:pass_rates}
\begin{tabular}{lcc}
\toprule
\textbf{Version} & \textbf{iOS} & \textbf{Android} \\
\midrule
V1 (Embedding) & 80.0\% & 82.0\% \\
V2 (Intent-based) & 91.6\% & 92.2\% \\
\bottomrule
\end{tabular}
\end{table}

The improvement comes from V2's ability to handle complex scenarios that defeated V1. Flows with stacked screens (like checkout processes with multiple forms), flows requiring repeated actions (like Uber Pool configuration), and flows with conditional paths based on user state all show dramatic improvements.

Breaking down by flow complexity:

\begin{itemize}
\item Simple flows (single screen, 3--5 actions): V1 80\%, V2 95\%
\item Medium flows (multiple screens, 10--15 actions): V1 75\%, V2 94\%
\item Complex flows (conditional branching, different actions on same screen, 20+ actions): V1 0\%, V2 89\%
\end{itemize}

V2 handles complexity that V1 could not address. The ability to maintain full action history and reason about screen context prevents the infinite loops and confusion that plagued V1 on complex flows.

We categorize non-passing runs into three groups:
\begin{itemize}
\item \textbf{Infrastructure failures} (\textasciitilde 1\% of runs): emulator, network, or device-farm issues, retried rather than blocked.
\item \textbf{Agent action errors} (\textasciitilde 0.5\%): the model took an ineffective action, caught by our error-handling mechanisms and retried.
\item \textbf{Genuine regressions} (the remainder): real defects from ongoing incidents or code changes, which the gate is designed to catch.
\end{itemize}

Pass rates also track flow priority. On the 61 core flows, the highest-priority journeys such as requesting and completing a trip, V2 sustains a near 99\% pass rate, with only infrastructure errors causing failures. Lower-priority flows fail more often and account for most of the gap to the aggregate, both because they exercise less-hardened paths and because their low-priority assertions are permitted to surface real but minor issues that, by policy, are tracked without blocking the commit.

\subsection{Comparison with External Systems}
\label{sec:external}

A same-benchmark comparison against prior LLM-based mobile testing systems is not available, because those systems target single-shot task automation or time-budgeted exploration rather than per-commit regression. To still provide an external reference point, Table~\ref{tab:external} collects the task-completion rates these systems report. Completion ranges widely, from 1.3\% for Guardian~\cite{guardian} and 39.4\% for DroidBot-GPT to 71.3\% for AutoDroid and 88.0\% for VisiDroid, the strongest baseline.

\begin{table}[htbp]
\centering
\caption{Task-completion rates for LLM-based mobile testing tools reported by VisiDroid~\cite{visidroid}.}
\label{tab:external}
\begin{tabular}{lcc}
\toprule
\textbf{Tool} & \textbf{Completion} & \textbf{Setting} \\
\midrule
Guardian~\cite{guardian}     & 1.3\%  & task automation \\
% DroidAgent~\cite{droidagent} & 24\%   & task automation \\
DroidBot-GPT~\cite{droidbotgpt} & 39.4\% & task automation \\
AutoDroid~\cite{autodroid}     & 71.3\% & task automation \\
VisiDroid~\cite{visidroid}     & 88.0\% & task automation \\
\midrule
DragonCrawl V2 (ours)          & 91.6 / 92.2\% & E2E regression gate \\
\bottomrule
\end{tabular}
\end{table}

These numbers are context, not a like-for-like result. The baseline figures are measured on small single-app task benchmarks under one attempt per task without the requirement to reproduce the same verdict on every code change. DragonCrawl's 91.6\% (iOS) and 92.2\% (Android) are measured on 1{,}013 multi-screen production flows run as a quality gates. The distinction that matters is therefore not the headline percentage, where VisiDroid at 88\% is close, but the setting: reliable, asserted, backend-aware flow completion repeated on every code change, which none of these systems was built to deliver.

\subsection{Latency Analysis}

Table~\ref{tab:latency} compares execution latency between V1 and V2. While individual LLM calls take approximately 0.5 seconds longer with GPT-4o compared to on-host MPNet inference, overall test execution is faster with V2 due to higher first-attempt pass rates.

\begin{table}[htbp]
\centering
\caption{Average Test Execution Latency.}
\label{tab:latency}
\begin{tabular}{lccc}
\toprule
\textbf{Flow Type} & \textbf{V1 (MPNet)} & \textbf{V2 (GPT-4o)} & \textbf{Delta} \\
\midrule
Simple flows & 5\,min & 6\,min & +1\,min \\
Medium flows & 16\,min & 13\,min & $-$3\,min \\
Complex flows & N/A & 20\,min & N/A \\
\bottomrule
\end{tabular}
\end{table}

The counterintuitive result occurs because V1's low per-call latency is offset by retry overhead. When V1 makes incorrect action predictions (especially on complex flows), tests must retry from checkpoints or restart entirely. V2's higher per-call latency (approximately 2.0\,s vs.\ 1.5\,s for V1) is more than compensated by getting actions right on the first attempt.

For complex flows, the improvement is dramatic. V1 had a 0\% pass rate on complex flows even after 3 retries. V2's 89\% pass rate means most tests succeed on the first attempt, reducing total execution time despite slower individual inferences.

This trade-off validates the decision to use more sophisticated (and slower) models: test reliability matters more than per-action latency in CI/CD environments where failed tests block deployments and require manual investigation.

% NEW CHANGE: new Ablation Study subsection addressing reviewer 2A ask for component-level ablations
\subsection{Ablation Study}

To quantify what each of DragonCrawl V2's design choices contributes, we ablated one component at a time and left the rest of the pipeline unchanged. We focused on the two components that most directly shape the agent's reasoning: prompt constraints, which enforce structured output, a valid action set, and execution guardrails; and view hierarchy canonization, which compresses the raw UI hierarchy into a compact, action-oriented representation before it reaches the LLM.

We evaluated each configuration on all the flows from the DragonCrawl repository. Because LLM agents are nondeterministic, we ran every flow five times under each configuration, for 5065 executions per configuration. Our primary metric was end-to-end success rate: a run counted as successful only if the agent satisfied every required mid-state assertion and reached the expected final state. We also reported the average number of actions and average execution time as efficiency measures.

\begin{table}[t]
\centering
\caption{Ablation study of DragonCrawl V2.}
\label{tab:ablation}
\setlength{\tabcolsep}{4pt}% default is 6pt
\begin{tabular}{lccc}
\toprule
Configuration & Success (\%) & Actions & Time (min) \\
\midrule
Full System & 95.9 & 15.1 & 4.8 \\
w/o Prompt Constraints & 89.6 & 16.4 & 5.7 \\
w/o View Canonization & 91.7 & 15.6 & 5.6 \\
\bottomrule
\end{tabular}
\end{table}

Table~\ref{tab:ablation} shows the results. Removing either component degraded quality, so both earn their place, but they helped in different ways.

\textbf{Prompt constraints} mattered most. Removing them dropped success from 95.9\% to 89.6\%. Without structured formats and an enforced action set, the agent more often emitted invalid actions or drifted from the intended flow, and recovering from those missteps lengthened trajectories, which raised both action count and time.

\textbf{View hierarchy canonization} improved efficiency. Removing it lowered success to 91.7\% and raised execution cost, because feeding the raw hierarchy buried the actionable elements in noise, so the agent spent more actions locating targets before it could act.

The two are therefore complementary: prompt constraints make each decision more reliable, while canonization makes reasoning cheaper by handing the model a compact, task-relevant view.

\textbf{Backend tool calling} is different in kind: it is not a knob that tunes reasoning but a prerequisite for authoring certain flows at all. Of 1{,}013 flows in scope, 234 (23.1\%) require at least one in-flow backend tool call, and without tool calling those flows cannot be authored or executed. We therefore treat it as an enabling capability rather than a component to ablate.

\textbf{A note on comparability:} the ablation figures should not be read against the deployment pass rates in Section~\ref{sec:pass_rate_analysis}. The ablation was run as a controlled study over a two-week window against a fixed app build, so no incoming code changes could introduce genuine regressions during the study; failures therefore reflect only agent behavior and residual infrastructure noise. The production rates in Table~\ref{tab:pass_rates}, by contrast, aggregate months of per-commit CI executions and include the regressions the gate exists to catch, along with ongoing incidents and week-to-week environment variability. The Full System row is thus the ablation's internal baseline for measuring component contributions, not a restatement of deployment performance, and the same applies to the execution times, which reflect the controlled environment.

\subsection{Cost Optimization Through Caching}

Initial cost estimates assumed naive LLM usage: full context re-sent on every inference call, all tests running on every commit. This projected to \$1.8M annually. While V1's on-host MPNet inference had virtually zero cost, the move to vendor LLMs was justified by observed industry trends showing token costs declining approximately 10$\times$ every 6--12 months, making the long-term cost trajectory favorable.

% TODO: Dhruva to fill out. We need a brief discussion that will then lead us to a table

At the optimized cost level, the system provides strong ROI. Compared to the estimated 27 developer years saved, the \$200K inference cost is highly cost-effective. The cost trade-off was clear: V1's zero inference cost could not offset the engineering hours lost maintaining Golden Datasets and the inability to test complex flows.

% \subsection{Developer-Year Saving Estimation}

% Developer-year savings are estimated as the sum of the one-time test authoring savings and the expected maintenance savings.

% The historical Dragon ROI calculation is:

% \[
% \text{Developer-years saved}
% =
% \frac{N_{\mathrm{tests}} \times T_{\mathrm{create}}}{H_{\mathrm{year}}}
% +
% \frac{N_{\mathrm{tests}} \times R_{\mathrm{failure}} \times T_{\mathrm{maint}}}{H_{\mathrm{year}}},
% \]

% where

% \begin{itemize}
% \item $N_{\mathrm{tests}}$ is the number of tests authored by Dragon;
% \item $T_{\mathrm{create}} = 16$ developer-hours, assuming manual authoring requires two workdays (one day for Android and one day for iOS);
% \item $R_{\mathrm{failure}} = 0.0035$ is the observed product-flow failure rate;
% \item $T_{\mathrm{maint}} = 16$ developer-hours, assuming each maintenance event requires approximately two workdays to resolve;
% \item $H_{\mathrm{year}} = 1380$ developer-hours per developer-year.
% \end{itemize}

% Substituting the constants gives

% \[
% \text{Developer-years saved}
% =
% \frac{N_{\mathrm{tests}} \times 16}{1380}
% +
% \frac{N_{\mathrm{tests}} \times 0.0035 \times 16}{1380}.
% \]

\section{Operational Experience}
\label{sec:operations}

Deploying DragonCrawl at production scale revealed insights beyond technical performance metrics. This section discusses operational efficiency gains, organizational transformation, adoption challenges, and practical lessons learned.

\subsection{Onboarding Efficiency}

Table~\ref{tab:onboarding} reports the distribution of onboarding time per flow. Under V1, engineers built a Golden Dataset for each flow by hand, capturing reference screens, labeling actions, and validating sequences. The median flow took about four days, with the 75th and 90th percentiles both at five days. The V1 sample was small and limited to low-complexity flows, so we do not read meaning into its tail; some flows simply took somewhat longer, and that is expected.

\begin{table}[htbp]
\centering
\caption{Onboarding Time per Flow and Flows Authored, V1 vs.\ V2.}
\label{tab:onboarding}
\begin{tabular}{lcccc}
\toprule
\textbf{Version} & \textbf{Median} & \textbf{p75} & \textbf{p90} & \textbf{Flows} \\
\midrule
V1 (Golden Dataset) & 4\,days  & 5\,days & 5\,days & 48 \\
V2 (Intent-based)   & 2.5\,hrs & 3\,hrs  & 4\,hrs  & 1{,}013 \\
\bottomrule
\end{tabular}
\end{table}

V2 replaced dataset construction with writing an intent definition: engineers state the goal in natural language, list mid-state assertions, and optionally supply tool call templates for backend setup, and the model handles navigation without reference data. The median flow took 2.5 hours, with the 75th and 90th percentiles at 3 and 4 hours. The spread here tracks flow complexity, as discussed above: simple flows are fastest to author, while flows with conditional paths or backend setup take longer.

At the median, this is a 30 to 40$\times$ reduction in onboarding time, and it is what enabled massive scale expansion. The team onboarded over 1{,}000 tests in the time it would have taken to manually annotate roughly 30 under V1.

% \subsection{Onboarding Efficiency}

% Table~\ref{tab:onboarding} compares test onboarding metrics. V1 required building extensive Golden Datasets. Engineers manually captured reference screens, labeled actions, and validated sequences. This took 4--5 days per flow.

% \begin{table}[htbp]
% \centering
% \caption{Test Onboarding Comparison.}
% \label{tab:onboarding}
% \begin{tabular}{lcc}
% \toprule
% \textbf{Metric} & \textbf{V1} & \textbf{V2} \\
% \midrule
% Time per flow & 96--120\,hrs & $<$4\,hrs \\
% Tests authored & 48 & 1013 \\
% \bottomrule
% \end{tabular}
% \end{table}

% V2 requires only writing intent definitions. Engineers specify the test goal in natural language, list mid-state assertions, and optionally provide tool call templates for backend setup. This takes 2--4 hours depending on flow complexity. The model handles navigation automatically without reference data.

% This 25--30$\times$ reduction in onboarding time enabled massive scale expansion. The team onboarded over 1,000 tests in the time it would have taken to manually annotate 40 tests under V1.

\subsection{Coverage Growth Over Time}

\begin{figure}[htbp]
\centering
\includegraphics[width=0.85\columnwidth]{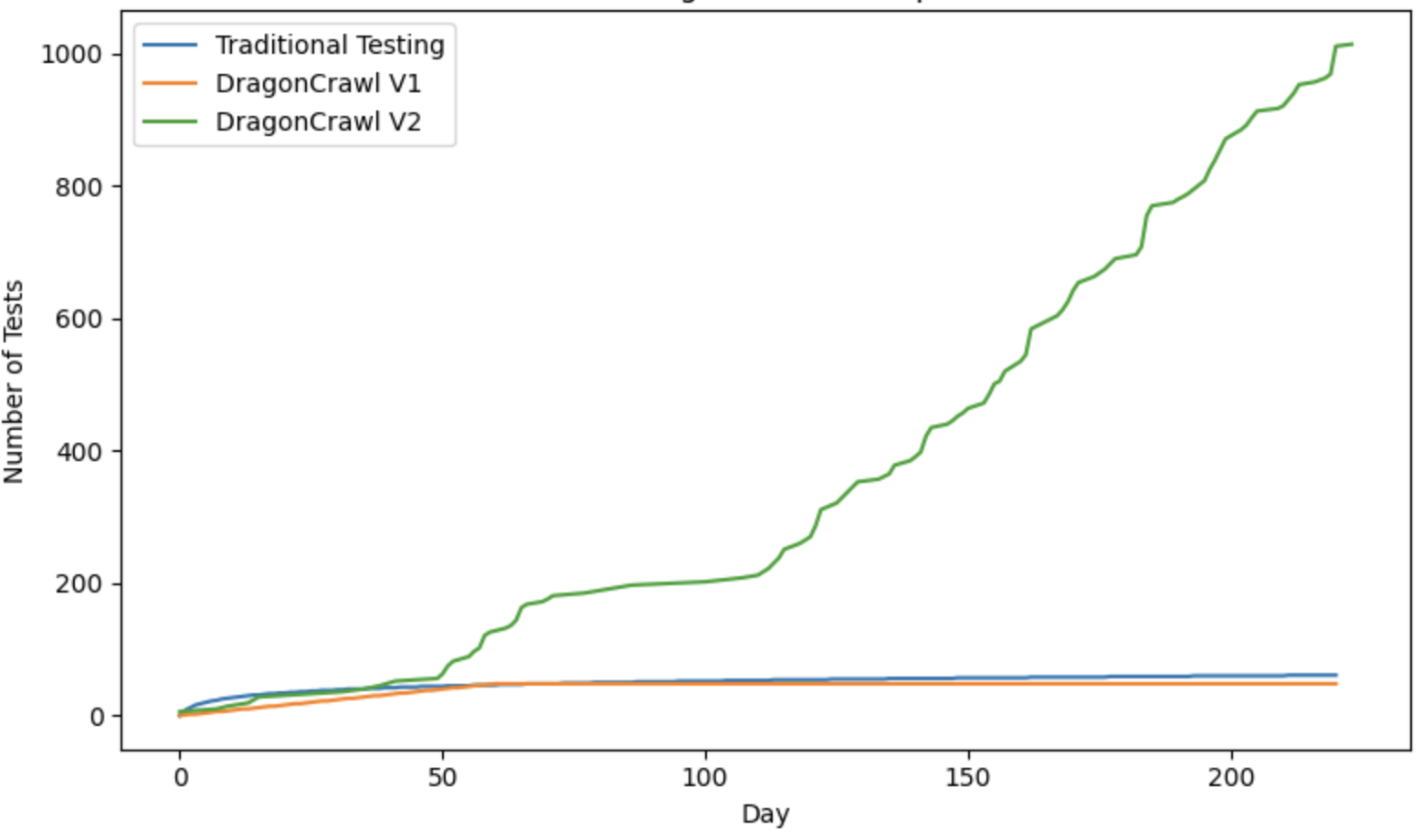}
\caption{Coverage growth over time.}
\Description{Line graph comparing test coverage growth within 200 days with traditional testing (showing flat growth), DragonCrawl V1 (showing logarithmic growth), and DragonCrawl V2 (showing pseudo-exponential growth)}
\label{fig:coverage_growth}
\end{figure}

Figure~\ref{fig:coverage_growth} shows test coverage growth over time. Pre-DragonCrawl, manual test creation grew logarithmically, and reached 61. Engineers could only add tests as fast as they could write and maintain them, and hence coverage plateaued. Then, V1 enabled linear growth and reduced maintenance overhead freed time for test creation. However, complex flow limitations and annotation costs constrained growth rate. Finally, V2 shows steeper linear growth approaching exponential growth because it significantly eased onboarding.

\subsection{Maintenance Overhead Reduction}

Prior to DragonCrawl, mobile engineers spent an estimated 30--40\% of their time on test maintenance. UI changes required updating test scripts. Flaky tests required investigation and fixing. Test infrastructure required constant attention.

Post-DragonCrawl, maintenance overhead dropped to approximately 5\%. The remaining maintenance involves:

\begin{itemize}
\item Updating intent definitions when test goals fundamentally change (rare)
\item Adjusting mid-state assertions when success criteria evolve
\item Adding new tool call templates for backend dependencies
\end{itemize}

Engineers cite this as the primary driver of job satisfaction improvements. Previously, significant time went to ``fixing'' tests that were not actually broken, just brittle. Now, engineers focus on building new features. When tests fail, it almost always indicates a real regression rather than test fragility.

\subsection{Estimated Developer-Year Savings}
We estimate savings against a traditional manual baseline in which each flow is scripted separately for Android and iOS, roughly 16 developer-hours per test, assuming $H_{\text{year}} = 1{,}380$ productive developer-hours per year.

\textbf{Authoring.} Through DragonCrawl, SDETs and mobile engineers have authored $N = 1{,}013$ tests from intent definitions. Scripting these by hand would cost $1{,}013 \times 16 = 16{,}208$ hours, which is roughly 12 developer-years of one-time authoring effort.

\textbf{Maintenance.} Manual scripts are brittle and need recurring upkeep as the UI evolves. Sustaining an equivalent manual suite over the deployment period would have required an estimated 20 hours of maintenance per test, or $1{,}013 \times 20 = 20{,}260$ hours, which is roughly 15 developer-years.

Combined, this is approximately 27 developer-years of engineering effort avoided. This uses ordinary manual scripting (16 hours per test) as the baseline rather than V1's far costlier Golden Datasets, so it is a conservative estimate.
% DragonCrawl has authored 1{,}013 tests that would otherwise be scripted and maintained by hand. Manual authoring takes about 16 developer-hours per test, one day each for Android and iOS, so writing all 1{,}013 by hand would consume roughly 12 developer-years ($1{,}013 \times 16 / 1{,}380$, at 1{,}380 productive hours per year). Keeping that suite alive as the UI evolves costs far more over time: at about 20 maintenance hours per test across the deployment period, maintenance alone accounts for another 15 developer-years. Together, DragonCrawl has saved an estimated 27 developer-years.

\subsection{Impact on Engineering Teams}

The transition to DragonCrawl fundamentally changed how mobile engineers approach quality assurance. Prior to the system, testing was viewed as a necessary burden. Engineers allocated significant time to test maintenance, accepting this as an unavoidable cost of quality.

Post-DragonCrawl, engineers report that testing feels like ``building'' rather than ``fixing''. The elimination of brittle locators means UI changes no longer cascade into test failures. Feature flags that modify screens do not require test updates. A/B experiments run without creating test maintenance debt.

This shift manifested in quantifiable productivity gains:

\begin{itemize}
\item Feature velocity increased by approximately 15\% as engineers spent less time on test maintenance
\item Time-to-market for new features improved as tests no longer bottlenecked releases
\item Bug escape rate to production decreased as engineers had time to write more comprehensive tests
\end{itemize}

Qualitative feedback from engineers emphasizes the ``Dragon\-Crawl Signal'' effect. Previously, failed builds were often ignored as ``probably just flaky tests''. Engineers developed learned helplessness around test failures, treating them as noise rather than signal. Post-DragonCrawl, red builds almost always indicate legitimate regressions. This clarity reduces developer burnout and improves response times to real issues.

Global autonomy represents another major benefit. Engineers can now launch features in new markets without waiting for localized QA teams to manually verify flows in specific languages. DragonCrawl handles multiple languages through the LLM's multilingual training, eliminating per-language test scripts. This accelerated international expansion and reduced coordination overhead.

\subsection{SDET Team Transformation}

The Software Development Engineer in Test (SDET) team, responsible for manual testing and QA, initially presented significant organizational resistance to DragonCrawl. The challenge was not technical but cultural.

\textbf{Initial Resistance}: SDET teams were accustomed to deterministic, script-driven testing workflows. DragonCrawl's autonomous navigation felt like erosion of their ``gatekeeper'' role. The move from manual execution to AI-driven testing raised concerns about job security and skill obsolescence. Many team members viewed the system as a threat rather than a tool.

\textbf{The Pivot}: The narrative shifted when the SDET team was repositioned as ``Context Engineers'' rather than ``Clickers''. Instead of executing repetitive test scripts, SDETs now build the knowledge base that guides DragonCrawl. They define user personas, specify environmental parameters, and create Model Context Protocols (MCPs) that inform the AI's decision-making.

This transformation elevated the role from manual execution to strategic orchestration, with SDETs performing the following:

\begin{itemize}
\item Managing agentic workflows, prompting the AI to find adversarial paths like ``Try to break the payment flow using an expired card in a low-connectivity environment''
\item Analyzing LLM logs to understand decision patterns and optimize prompts
\item Developing tool call templates to enable complex  scenarios
\item Contributing to the Golden Datasets by reviewing and validating novel interaction sequences
\end{itemize}

\textbf{Upskilling Impact}: The transition required significant training investment. SDET members learned prompt engineering, basic Python scripting for test definition, and how to interpret LLM behavior logs. This upskilling has high market value, transforming ``manual tester'' roles into ``QA AI Engineer'' positions. Team members report increased job satisfaction and career advancement opportunities.

The organizational re-leveling solved the ``Scalability Paradox'', the observation that as applications grow, traditional approaches require exponentially more people for testing. By offloading the 30\% maintenance tax from mobile engineers and elevating SDETs from manual execution to AI orchestration, Uber achieved scalable quality assurance without proportional headcount growth.

\subsection{Bug Detection and Resolution}

DragonCrawl has detected numerous high-severity issues that would have reached production under previous testing approaches. Notable examples include:

\textbf{Internationalization Bugs}: The system automatically tests in 20+ languages, catching unlocalized strings and text truncation issues that manual testing missed. Over 100 i18n bugs were identified in the first six months.

\textbf{Payment Flow Issues}: Complex payment scenarios (expired cards, insufficient balance, promotional code interactions) revealed edge cases where error handling was incomplete. DragonCrawl's tool calling enabled setting up these scenarios systematically.

\textbf{State Management Bugs}: Several flows failed when users navigated back and forth between screens. These were invisible to tests that followed linear paths but were caught by DragonCrawl's exploratory behavior within goal constraints.

\begin{figure}[htbp]
\centering
\includegraphics[width=\columnwidth]{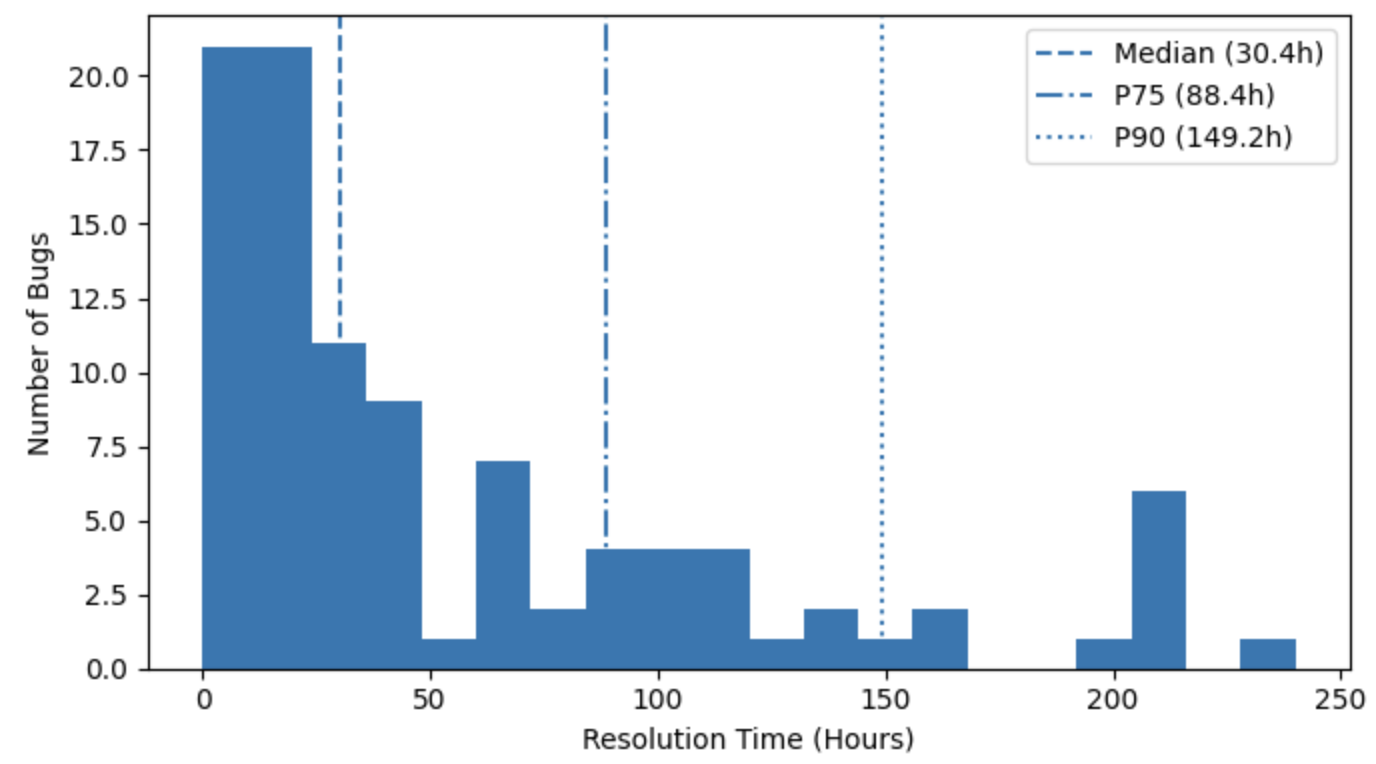}
\caption{Distribution of high priority bug resolutions.}
\Description{Histogram showing distribution of high priority bug resolution times in hours, with median at 30.4 hours, p75 at 88.4 hours, and p90 at 149.2 hours}
\label{fig:bug_resolutions}
\end{figure}

Finally, bug resolution time and reassignment rates improved substantially with DragonCrawl. Figure~\ref{fig:bug_resolutions} shows the distribution of high-priority bug resolution times. The median resolution time is 30.4 hours (1.3 days), compared to approximately 5 days under traditional testing. The primary driver of this improvement is automation in triage and ownership. Under traditional testing, bug reports required manual triage and manual reassignment across teams, with a median of three reassignments per high-priority issue. In contrast, DragonCrawl executes on every PR and automatically files tickets directly to the PR owner. Across all 98 high-priority bugs detected by DragonCrawl, we observe zero reassignments.

While the median resolution time is 1.3 days, the 75th percentile is 3.7 days and the 90th percentile is 6.2 days. The heavier tail is largely explained by temporal effects: issues introduced late on Fridays are typically resolved the following Monday, and resolution times can increase further during holidays or temporary code freezes.

\subsection{Adoption Challenges}

Despite overall success, several challenges emerged during rollout:

\textbf{Trust Building}: Engineers initially distrusted non-deterministic testing. The concern was that AI-driven tests might pass incorrectly, missing real bugs. Building trust required transparency: showing decision logs, explaining why the model took specific actions, and demonstrating high correlation between DragonCrawl results and manual verification.

\textbf{Learning Curve}: Writing effective intent definitions and mid-state assertions required understanding what the LLM could and could not do. Engineers needed training on prompt engineering principles and common pitfalls. Early test definitions were often too vague or overly specific, requiring iteration to find the right level of abstraction.

\textbf{Infrastructure Stability}: Integrating LLM inference into CI pipelines introduced new failure modes. Network timeouts to external API endpoints, rate limiting, and occasional model unavailability required robust retry logic and fallback mechanisms.

\textbf{Cost Visibility}: Engineering teams needed education on cost implications of test design. Excessive mid-state assertions or overly complex flows could drive up inference costs. Dashboards showing per-test costs helped teams optimize while maintaining coverage.

\subsection{Lessons Learned}

Several key lessons emerged from production deployment:

\textbf{Start Simple}: Initial rollout focused on well-understood flows with clear success criteria. This built confidence and allowed iterative improvement before tackling complex scenarios. Attempting to solve all problems simultaneously would have failed.

\textbf{Invest in Observability}: Comprehensive logging of LLM decisions, screen states, and action sequences was essential for debugging and optimization. Without visibility into the model's reasoning, failures would be opaque and trust would not be built.

\textbf{Human-in-the-Loop Validation}: While automation is the goal, human review of novel interaction sequences maintains quality. Periodic sampling of test executions for manual verification catches edge cases and provides feedback for prompt refinement.

\textbf{Cost Optimization is Critical}: LLM inference costs can spiral without careful engineering. Caching, selective execution, and batch processing are not optional optimizations but core requirements for production deployment.

\textbf{Organizational Change Requires Attention}: Technology alone is insufficient. Addressing concerns from teams whose work is transformed, providing training and support, and communicating vision clearly are as important as technical implementation.

\section{Related Work}
\label{sec:related}

Early mobile testing tools like Monkey~\cite{monkey} used random input generation, achieving low coverage. Model-based approaches like Sapienz~\cite{sapienz}, later deployed at scale at Facebook~\cite{metasapienz}, and Stoat~\cite{stoat} improved coverage with genetic algorithms and stochastic models but were brittle when the UI changed, with fragility and UI evolution consuming around 30\% of test maintenance~\cite{coppola2020}. Learning-based approaches like Humanoid~\cite{humanoid}, DroidBot~\cite{droidbot}, and Q-Testing~\cite{qtesting} added adaptivity but still operated on low-level UI elements without high-level goal understanding.

Recent work applies LLMs to mobile testing: GPTDroid~\cite{gptdroid} reports 71\% activity coverage, AutoDroid~\cite{autodroid} and DroidAgent~\cite{droidagent} use LLM-based planning, and VisiDroid~\cite{visidroid} adds visual reasoning for task-completion detection. Complementary work targets UI display defects~\cite{owleyes, ranvisual}, context-aware text input generation~\cite{fillblank}, and change-targeted testing~\cite{hawkeye}. The LLM-based exploratory systems measure success by activity coverage and crashes found over fixed exploration periods. Where they report goal completion, DragonCrawl reaches parity with or exceeds the strongest of them, including VisiDroid (Section~\ref{sec:external}); we treat this as a reference point, not a shared benchmark, since those rates come from small single-app tasks under a single attempt.

% Recent work applies LLMs to mobile testing: GPTDroid~\cite{gptdroid} reports 71\% activity coverage, AutoDroid~\cite{autodroid} and DroidAgent~\cite{droidagent} use LLM-based planning, and VisiDroid~\cite{visidroid} adds visual reasoning for task-completion detection. Complementary lines of work target specific aspects of the problem: visual understanding for UI display defects~\cite{owleyes} with industrial deployment~\cite{ranvisual}, context-aware text input generation~\cite{fillblank}, change-targeted test selection via reinforcement learning~\cite{hawkeye}, test generation from usage videos~\cite{avgust}, and LLM-based unit test generation~\cite{testpilot}. The LLM-based exploratory systems measure success by activity coverage and crashes found over fixed exploration periods. Where they report goal completion, DragonCrawl reaches parity with or exceeds the strongest of them, including VisiDroid (Section~\ref{sec:external}); we treat this as a reference point, not a shared benchmark, since those rates come from small single-app tasks under a single attempt.

The fundamental difference is setting: exploration asks ``What can we discover?'' and runs occasionally, whereas regression asks ``Does this change break existing functionality?'' and runs on every commit. DragonCrawl validates 61 critical flows per code change in CI/CD, completing specific journeys across markets and languages and blocking commits that break them. This demands reliable, repeatable completion of the same flow rather than the non-determinism exploratory tools tolerate. Combining intent-based reasoning, multimodal assertion, backend integration through tool calling, and a production deployment of 1{,}013 tests, DragonCrawl is, to our knowledge, the first LLM-based mobile testing system operating as a per-commit regression gate at this scale.

\section{Limitations}
\label{sec:limitations}

Despite significant successes, DragonCrawl V2 has several limitations that organizations should consider before adoption.

\textbf{External Dependencies and Coverage Gaps}: DragonCrawl relies on GPT-4o, an external commercial model. This creates risks around model updates that may affect behavior, pricing changes that could impact ROI, and availability concerns during outages. Organizations with strict data residency requirements may not be able to use external LLM services and would need to explore self-hosted alternatives.

\textbf{Cost and Scale Barriers}: Even after aggressive optimization, inference costs roughly \$200K annually, a budget few organizations can dedicate to testing tokens alone. The ROI case also depends on Uber's scale: with a smaller test suite, the developer-years saved may not offset inference costs, engineering investment, and the evaluation infrastructure required. Similarly, our continuous evaluation relies on a curated dataset of roughly 10,000 validated trajectories accumulated over years of operation; organizations starting fresh cannot easily replicate this safety net.

\textbf{Residual Non-Determinism and Assertion Reliability}: Despite constraint-based prompting, the agent occasionally takes valid but unintended paths, which can mask bugs on the specific path a test owner meant to exercise. Visual assertions, while robust to UI drift, inherit the failure modes of the underlying model: transient states such as loading screens can produce false verdicts, and low-priority assertion mismatches are by policy non-blocking, so some real but minor defects reach production by design.

\textbf{Generalizability}: Our results come from a single company, two first-party applications, and one model family. DragonCrawl benefits from Uber-specific infrastructure, including controlled test environments, backend tool templates, and mature CI/CD, that other organizations would need to build first. We have not evaluated the approach on third-party apps, other app categories such as games or canvas-heavy UIs where view hierarchies are uninformative, or smaller models, so the reported pass rates may not transfer.

\section{Conclusion}
\label{sec:conclusion}

DragonCrawl V2 demonstrates that intent-based AI-driven testing can address fundamental limitations of traditional mobile testing approaches. By leveraging generative reasoning with large language models, the system eliminates brittle element locators, reduces test maintenance overhead from 30--40\% to approximately 5\% of engineering time, and enables comprehensive coverage across languages, markets, and flow variations without per-variant test scripts.

A key contribution is the shift from exploratory testing to continuous regression testing. Prior LLM-based testing research focused on maximizing coverage and discovering crashes. DragonCrawl instead validates specific user flows on every code change, blocking commits that break critical functionality. The system runs 1,013 tests in CI/CD pipelines, providing go/no-go signals for deployments rather than periodic bug reports.

Production deployment at Uber validates the approach at scale. With 91.6\% pass rate on iOS and 92.2\% on Android across 1,013 automated tests, DragonCrawl has saved an estimated 27 developer years in authorship and maintenance efforts while enabling test coverage expansion that would be impossible with traditional approaches. The organizational transformation deserves emphasis: SDET teams evolved from manual test executors to AI orchestrators, while engineers regained confidence in test signals as DragonCrawl's robustness eliminated flaky test problems.

Key lessons include starting with simple flows before tackling complex scenarios, investing in observability and logging, maintaining human-in-the-loop validation, and treating cost optimization as a first-class concern. The transition from similarity-based matching to generative reasoning represents a fundamental shift: rather than encoding specific interaction sequences, we specify what we want to accomplish and let the model reason about how to achieve it. This paradigm change makes testing robust to UI evolution, scalable across dimensions of variation, and maintainable over time.

\section*{Data Availability Statement}

DragonCrawl is proprietary to Uber Technologies, Inc. Its code, prompts, models, and evaluation data are tied to internal infrastructure and business-sensitive logs from the Uber apps and therefore cannot be released. To support replication, we provide detailed architecture, prompt strategies, constraints, and quantitative results.

\begin{acks}
The authors thank the Uber Mobile Engineering and SDET teams for their contributions to DragonCrawl development and deployment. We thank the Mobile Foundations, Continuous Integration, and ML Platform teams for infrastructure support. 
\end{acks}

\balance
\bibliographystyle{ACM-Reference-Format}
\bibliography{references}

\end{document}